\documentclass[aps,twocolumn,%
showkeys,showpacs,nofootinbib,byrevtex]{revtex4-1}
\usepackage{amsmath,amsfonts,amssymb,amscd,amsxtra,amsthm}
\usepackage{graphicx}
\usepackage{epsfig,bm,dcolumn}
\usepackage{epstopdf}
\usepackage{xcolor}

\begin{document}


\title{Photoproductions of $f_1(1285)$ and $\eta'(958)$
from the analysis of CLAS data with the Primakoff effect at high
energies}


\author{Byung-Geel Yu}%
\email{bgyu@kau.ac.kr}

\author{Kook-Jin Kong}%
\email{kong@kau.ac.kr}

\affiliation{Research Institute of Basic Sciences, Korea Aerospace
University, Goyang 10540, Korea}

\date{\today}


\begin{abstract}
We investigate photoproduction of axial vector meson $f_1(1285)$
based on the CLAS experiment $\gamma p\to pf_1(1285)\to
p\eta\pi^+\pi^-$ using the Regge model for $\rho^0+\omega$
exchanges. The combined analysis of $\gamma p\to p\eta'\to
p\eta\pi^+\pi^-$ reaction including $\eta(1295)$ is accompanied to
evaluate the decay mode $\eta(1295)\to \eta\pi^+\pi^-$ from the
potential overlap with the $f_1$ reaction. The predominance of
$\rho^0(770)$ exchange over the $\omega(782)$ with the coupling
constant $\gamma\rho^0f_1$ extracted from the CLAS experiment
leads to a good description of the measured cross sections for
$f_1$, while the scale of the cross section is corrected by the
branching fraction $\eta'\to\eta\pi^+\pi^-$, and
$f_1\to\eta\pi^+\pi^-$, respectively. As an extension to study the
nonmesonic production mechanism, the Primakoff effect by the
virtual photon exchange is investigated in the exclusive reaction
$\gamma p\to pf_1$ up to $\sqrt{s}\approx50$ GeV, finding that it
plays a similar role to the Pomeron in the vector meson
photoproduction at high energies. Such an unique role of the
virtual photon exchange similar to the $f_1$ case is identified up
to $\sqrt{s}\approx250$ GeV in the $\gamma p\to p\pi^0$ and
$\gamma p\to p\eta$ as well as $\gamma p\to p\eta'$ reaction.
\end{abstract}


\maketitle

\section{Introduction}

The axial vector meson $f_1(1285)$ of spin-isospin and
parity quantum numbers $I^GJ^{PC}=0^+(1^{++})$ has many
interesting aspects from the standpoint of QCD; the mixing
of gluon contents in the $f_1$ wave function associated
with $U(1)_A$ anomaly \cite{kochelev1} and its application
to vector meson photoproduction \cite{omega,phi},
the Primakoff effect via the $f_1\to\gamma^*\gamma$ decay
in the presence of electromagnetic interaction \cite{osipov},
and the branching ratio $f_1\to a^0 \pi\approx 36\,\%$
which is large enough to study the exotic four-quark state
$a_0(980)$ \cite{oset}. Moreover, the decay width
$\Gamma_{f_1}=22.7\pm1.1$ MeV from Particle Data Group (PDG)
(and 18.4$\pm$1.4 MeV from the recent measurement by the
CLAS Collaboration \cite{dickson}) is much smaller than the
typical meson decay width, which can be advantageous in
finding the formation of exotics such as $\pi_1(1400)$
via the $\gamma p\to\Delta\pi_1\to p\eta\pi\pi$ process
\cite{szcze,jpac1,ghoul,jpac2,kuhn,schott}.

Recent reports of differential cross sections for $\gamma p \to
pf_1(1285)$ measured from the reaction $\gamma p\to
p\eta\pi^+\pi^-$ by CLAS Collaboration \cite{dickson} draw our
attention, because it is the first measurement to provide
information on the static and dynamical properties of $f_1$ meson
that can access the above issues through theoretical studies and
comparisons of experiments on $f_1$ photoproduction. On the other
hand, however, the existing models that predict $f_1$
photoproduction \cite{kochelev,domokos} using the $\rho^0+\omega$
Regge poles in the $t$-channel are poor to agree with CLAS
experimental data. This is mainly because their contributions with
the anomaly coupling constants of $\gamma\rho^0 f_1$ (and
$\gamma\omega f_1\approx{1\over3}\gamma\rho^0 f_1$) determined
from PDG are too large to be consistent with data. Furthermore,
since the exclusive reaction $\gamma p\to pf_1$ should be
reconstructed through the multi-meson production in the final
state from the aforementioned CLAS data, experimental
circumstances such as the branching ratio $f_1\to\eta\pi^+\pi^-$
should be considered in the analysis of CLAS data for the single
process $\gamma p\to pf_1$.

In this work, we investigate photoproduction of $f_1$ based on the
CLAS data of the reaction $\gamma p\to p\eta\pi^+\pi^-$ using a
Reggeized model for $\rho^0+\omega$ exchanges. We recall that the
data were extracted from the experiment in which case the
interference of $\eta(1295)$ photoproduction was neglected in the
region overlapping with $f_1(1285)$. Viewed from the PDG,
productions of $\eta'$ and $\eta(1295)$ as well as $f_1(1285)$ are
other sources of decay to $\eta\pi^+\pi^-$, so the reactions
$\gamma p\to p\eta'$ and $\gamma p\to p\eta(1295)$ are also
involved in the reaction $\gamma p\to  p\eta\pi^+\pi^-$. Thus, we
examine the respective contributions of $\eta'$ and $\eta(1295)$
photoproductions to the CLAS data first, starting from
photoproductions of these $\eta$'s. The $f_1$ photoproduction will
then be described with the trajectories of $\rho^0+\omega$ Regge
poles and their $VNN$ coupling constants the same as in the case
of $\eta$ and $\eta'$. The reliability of model prediction is
confirmed by the roles of $\rho^0+\omega$ Regge poles in the
$\eta$ and $\eta'$ cases, which were rather well established in
previous studies \cite{chiang,tiator}.
In the model study of $f_1$ photoproduction, the $\rho^0$ exchange
plays the leading role over the $\omega$. Therefore, it is of
importance to consider the decay width $f_1\to\rho^0\gamma$ more
appropriate for a consistent description with CLAS data. As
mentioned earlier, the PDG value is very large and there is a
significant difference from the one extracted from the CLAS
experiment. In Ref. \cite{osipov} this issue was revisited to
evaluate the decay width $f_1\to\rho^0\gamma$ based on the
well-known triangle diagram for the $AVV$ coupling vertex. The
quark loop calculation using Bose symmetry and gauge invariance
yields the decay width $f_1\to\rho^0\gamma$ much smaller than the
PDG value, supporting the one extracted from the CLAS experiment
\cite{dickson}.

In addition to the vector meson exchange we consider another
subprocess of nonmesonic exchange that has different energy
dependence from vector mesons. By the $C$-even property of $f_1$
meson and $\eta$ as well, the virtual photon of $C$-odd is allowed
to exchange in the $t$-channel, the so-called the Primakoff effect
that manifests itself at very forward angles as a rapid
enhancement in the differential cross section
\cite{kaskulov,sibirtsev}. In the energy range of the CLAS
experiment where the vector meson exchanges are dominant the
Primakoff effect is expected to be suppressed by the charge
coupling with the nucleon. Nevertheless, since the exchange of
virtual photon is not to be Reggeized, its role could become
significant at high energy, where there is decrease of vector
meson exchange according to the energy dependence $\sim
s^{\alpha_V(0)-1}$.

Now that the Primakoff effect in  the photoproduction of $\eta$
and of $\eta'$ is related to the flavor mixing of $\eta$-$\eta'$
(the theme of the PrimEx project at CLAS \cite{gasparian} and the
current upgrade to GlueX at CLAS12), measurements of the Primakoff
effect in the reaction can give clues to understanding the
structure of the flavor symmetry, e.g., the mixing of flavor octet
and singlet between $\eta$-$\eta'$, and so is the mixing between
$f_1(1285)$-$f_1(1420)$ \cite{yang}. This point will be addressed
in the context of the two-gamma decay in the photoproduction of
pseudoscalar meson $\pi^0$, $\eta$ and $\eta'$, and will be
extended to the Primakoff effect in the $\gamma p\to pf_1$
reaction at high energies.

This paper is organized as follows: In Section II,
photoproductions of $\eta$, $\eta'$ and $\eta(1295)$ on proton
target are investigated in the Reggeized model for $\rho^0+\omega$
exchanges. Sec. III devotes to an analysis of the exclusive $f_1$
photoproduction on proton from the multimeson reaction $\gamma
p\to p\eta\pi^+\pi^-$ within the same approach as in Sec. II.
Differential and total cross sections for the CLAS experiment are
reproduced in the subsection (a). Predictions for the energy
dependence of differential cross section and the beam polarization
asymmetry are presented to distinguish between the reactions
aforementioned for future experiments. The Primakoff effect by the
virtual photon exchange is studied in the photoproduction of $f_1$
in the subsection (b), and $\pi^0$, $\eta$ and $\eta'$ cases in
(c). Summary and conclusion are given in the Section IV.

\section{photoproductions of $\eta(548)$, $\eta'(958)$, and
of $\eta(1295)$ on the proton target}

As the threshold energies of the exclusive processes $\gamma p\to
p\eta(1295)$ and $\gamma p\to p f_1(1285)$ are over the region
$\sqrt{s}_{thres.}\approx 2.2$ GeV, the contribution of nucleon
resonances in the direct and crossed channels are not expected, and,
hence, it is good to consider only the $t$-channel meson exchange
for the description of these reactions.

In this section we treat the photoproduction of $\eta'$ and
$\eta(1295)$ in a single framework where the vector meson exchange
is reggeized with vector meson nucleon coupling constants $(VNN)$
common in all the reactions we are dealing with in the current
work. For consistency, let us start from $\gamma p\to p\eta(548)$
to provide a basic formalism which will be extended to $\eta'$ and
$\eta(1295)$ with each radiative decay constant determined from
the empirical decay width for the different $\eta$ mass,
respectively.

\begin{figure}[]
\centering \epsfig{file=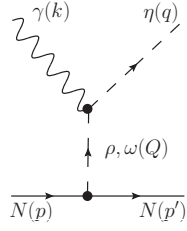, width=0.3\hsize}
\caption{$\rho$ and $\omega$ exchanges in the
exclusive $\eta$ photoproduction. $Q^\mu=(q-k)^\mu$
denotes the momentum transfer in the $t$-channel.}
\label{fig1}
\end{figure}

For the exclusive $\eta$ photoproduction on nucleon,
\begin{eqnarray}
\gamma(k)+N(p)\to \eta(q)+N(p')
\end{eqnarray}
we denote the particle momenta $k$, $q$, $p$ and $p'$ to stand for
the incident photon, outgoing $\eta$ meson, and the initial and
final nucleons, respectively. $s=(p+k)^2$ and $t=(q-k)^2$ are the
Mandelstam variables in the reaction kinematics. We restrict our
discussion only to the production mechanism by the meson exchange,
as depicted in Fig. \ref{fig1}, for our purpose here is to see how
the meson exchange works well in the kinematical region of
$\eta(1295)$ and $f_1(1285)$ photoproductions.

The photoproduction amplitude for the exchange of $C$-odd vector
meson on nucleon is written as,
\begin{eqnarray}
&&M(\gamma N\to N\eta)=\pm\rho+\omega,\label{amp1}
\end{eqnarray}
with the sign of $\rho$ for proton and neutron, respectively.

The effective Lagrangians for the vector meson exchange is written as
\begin{eqnarray}
&&{\cal L}_{V NN}=\bar{N}\left(g^v_{VNN}\gamma^\mu
+\frac{g^t_{VNN}}{2M}\sigma^{\mu\nu}\partial_\nu\right)V_\mu N
\label{vnn}\\%
&&{\cal L}_{\gamma\eta V}=\frac{g_{\gamma\eta
V}}{4m_0}\epsilon_{\mu\nu\alpha\beta}
F^{\mu\nu}V^{\alpha\beta}\eta+\mbox{H.c.},\label{etann}
\end{eqnarray}
and the corresponding production amplitude is given by
\begin{eqnarray}
\label{vec}
&&{\cal M}_{V}=\frac{g_{\gamma\eta V}}{m_0}
\epsilon_{\mu\nu\alpha\beta}\epsilon^{\mu}k^{\nu}Q^{\alpha}
\left(-g^{\beta\rho}+ {Q^{\beta}Q^{\rho}\over m_V^2}\right)
\nonumber\\&&
\times\overline{u}'(p')\left( g^v_{VNN}\,\gamma_{\rho} +
\frac{g^t_{VNN}}{4M}[\gamma_\rho, \rlap{\,/}{Q}]
\right)u(p){\cal R}^V(s,t).
\end{eqnarray}
Here, $\epsilon^\mu$ is the incident photon polarization and
$Q^\mu=(q-k)^\mu$ is the  momentum transfer in the $t$-channel
with $Q^2=t$. The coupling constant is normalized by
the mass parameter $m_0$ chosen to be 1 GeV.

The Regge pole for spin-1 vector meson is given by
\begin{eqnarray}\label{vec-regge}
{\cal R}^V(s,t)
=\frac{\pi\alpha'_V\times{\rm phase}}{\Gamma[\alpha_V(t)]
\sin\pi\alpha_V(t)}
\left(\frac{s}{s_0}\right)^{\alpha_V(t)-1}
\ \
\end{eqnarray}
for the vector meson $V(=\rho,\ \omega)$ collectively.

The choice of $\rho$ trajectory is not unanimous. Viewed from the
reactions \cite{yu-kong-rho,yu-kong-pin} where the single $\rho$
exchange is involved to test its role, it is better to choose
$\rho$ trajectory as
\begin{eqnarray}\label{trajectory}
&&\alpha_{\rho}(t)=0.9\,t+0.46,\\
&&\alpha_{\omega}(t)=0.9\,t+0.44,
\label{omega-traj}
\end{eqnarray}
and for the $\omega$ trajectory as well with the coupling constants
$g^v_{\rho NN}=2.6$ and $g^t_{\rho NN}=9.62$. For the
$\omega NN$ couplings, we use $g^v_{\omega NN}=15.6$ and
$g^t_{\omega NN}=0$ consistent with the ratio
$f_{\rho}:f_{\omega}=1:3$ by the vector meson dominance.

The radiative decay constant $g_{\gamma\eta V}$ is
determined from the measured decay width,
\begin{eqnarray}\label{eq:etapdecay}
 \Gamma_{V \rightarrow \eta\gamma}
 = \frac{1}{96\pi}\frac{g^2_{\gamma\eta V}}{m_0^2}
 \left(\frac{m_{V}^2-m_\eta^2}{m_V}\right)^3.
\end{eqnarray}
For the process $\gamma p\to p\eta'$ and $\gamma p\to p\eta(1295)$
in which cases the decay modes are reversed, i.e., $\eta'\ (\eta(1295))\to V$,
the decay width in Eq. (\ref{eq:etapdecay}) is multiplied by
the factor of 3 to recover the initial $\eta'\ (\eta(1295))$
spin degree of freedom.

\begin{table}\caption{Compilation of coupling constants
used for the eta-maid \cite{chiang} and
Regge model \cite{kochelev} for $\eta$ photoproduction.
Radiative coupling constant $g_{\gamma\eta V}$ is given
in units of GeV$^{-1}$.
}
\begin{ruledtabular}\label{tb1}
\begin{tabular}{lccccccl}
                       &  $\eta$-MAID \cite{chiang}& Regge model \cite{kochelev} & This work        &\\
\hline
$g_{\gamma\eta\rho}$   & $0.448$     &  -        & $0.448$    & \\%
$g_{\gamma\eta'\rho}$  &$0.392$      &  -        & $0.36$     & \\%
$g_{\gamma\eta(1295)\rho}$  &-       &$0.0566$  & $0.0566$   & \\%
$g^v_{\rho NN}$        &2.4          &3.9       &    2.6    & \\%
$g^t_{\rho NN}$        &8.88         &23.79     &    9.62   & \\%
\hline
$g_{\gamma\eta\omega}$ & $0.16$      &  -        & 0.106     & \\%
$g_{\gamma\eta'\omega}$& $-0.136$    &  -        & $0.12$    &  \\%
$g_{\gamma\eta(1295)\omega}$& -      &0.0189    & $0.0189$    &  \\%
$g^v_{\omega NN}$      & 9           &10.6      &   15.6     &       \\%
$g^t_{\omega NN}$      & 0           & 0        &  0          &       \\%
\end{tabular}
\end{ruledtabular}
\end{table}

Given the trajectories in Eqs. (\ref{trajectory}) and
(\ref{omega-traj}) together with the ratios
$g_{\gamma\eta\rho}/g_{\gamma\eta\omega}\simeq 4$ and $2g^v_{\rho
NN}/g^v_{\omega NN}\simeq1/3$ in Table \ref{tb1}, the $\omega$
exchange  is expected to give the contribution quite the same as
the $\rho$, if the same phase is taken. This observation could be
valid for $\eta'$ and $\eta(1295)$ as well within the present
framework. We choose the complex phase $e^{-i\pi\alpha_V(t)}$ for
both $\rho$ and $\omega$ Reggeons in the case of proton target,
because there is no dip from the nonsense zeros of the Regge poles
in the differential cross section data
\cite{braunschweig70,dewire,crede}. This agrees with the general
features of reaction cross sections. For the reaction $\gamma n\to
n\eta$, we take the constant phase $1$ for $\rho$, and $-1$ for
$\omega$ Reggeons for a better description of the total cross
section.

\begin{figure}[]
\centering \epsfig{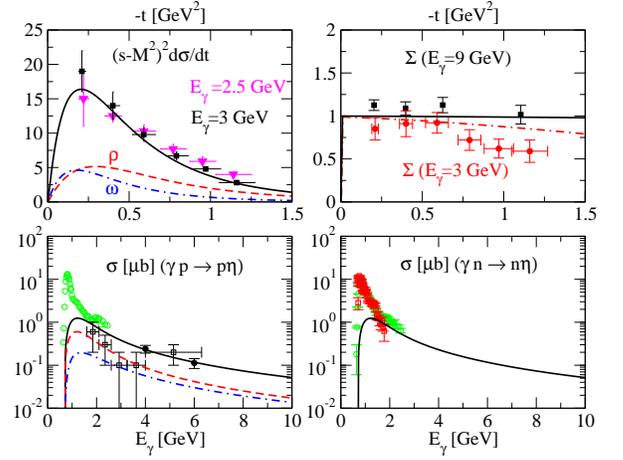} \caption{Scaled
differential cross section, beam polarization, total cross
sections for $\gamma p\to p\eta$ and total cross section for
$\gamma n\to n\eta$. The solid curve results from $\rho^0+\omega$
Reggeon exchanges. Scaled differential cross section at
$E_\gamma=3$ GeV is shown with data taken from Ref.
\cite{bussey76}, and for beam polarization with respect to $-t$ at
$E_\gamma=3$ GeV \cite{bussey76} and at 9 GeV \cite{ghoul}. Data
for $p\eta$ total cross section at $E_\gamma=4$ and 6 GeV are from
the integration of DESY differential data in Ref.
\cite{braunschweig70} and open squares in the $1.6\leq
E_\gamma\leq 6.3$ GeV are from AHHM data in Ref. \cite{struc76}.
Data for resonance peaks are from
world dataset \cite{jaegle,crede05,crede09}. \\
} \label{fig2}
\end{figure}

To illustrate the validity of the $\rho+\omega$ Reggeon exchanges
for the reaction $\gamma N\to N\eta$, we present the scaled
differential cross section by the factor $(s-M^2)^2$ reproduced at
$E_\gamma=3$ GeV(solid curve), beam polarizations at $E_\gamma=3$
GeV(dash-dotted) and $E_\gamma=9$ GeV(solid), and the respective
total cross sections for $\gamma p\to p\eta$ and $\gamma n\to
n\eta$ as well in Fig. \ref{fig2}. The (red) dashed and (blue)
dash-dotted curves correspond to the $\rho$ and $\omega$ Reggeon
contributions to differential and total cross sections from a
proton taget, respectively. As mentioned before, they play the
role roughly equal to each other. A few remarks are in order;
$\eta$ photoproduction is sensitive to a choice of phase of
Reggeon as well as the $\rho$ trajectory between
$\alpha_\rho=0.8t+0.55$ and that in Eq. (\ref{trajectory}). In the
former case, it is advantageous to choose one of the Reggeon,
i.e., $\omega$ to have the exchange-nondegenerate phase. But, in
that case, the contribution of the $\omega$ is suppressed, and the
production mechanism resulting from the dominance of $\rho$ over
the $\omega$ would be quite different from the present one as
shown in Fig. \ref{fig2}.

In the meanwhile, most of the Regge models for $\eta$
photoproduction introduce hadron form factors at the $\gamma\eta V$
and $VNN$ vertices to fit to
experimental data \cite{chiang}. However, it is natural to dispense with
such form factors in the Regge amplitude, because it
contains the gamma function $\Gamma(\alpha_V(t))$ in Eq.
(\ref{vec-regge}) to suppress the singularity from the sequential
zeros of $\sin\pi\alpha_V(t)$. Thus, as
demonstrated in Figs. \ref{fig5} and \ref{fig6}, the present approach
without form factors is less model dependent than those in Refs.
\cite{kochelev,domokos} to describe the CLAS data.

\begin{figure}[]
\centering \epsfig{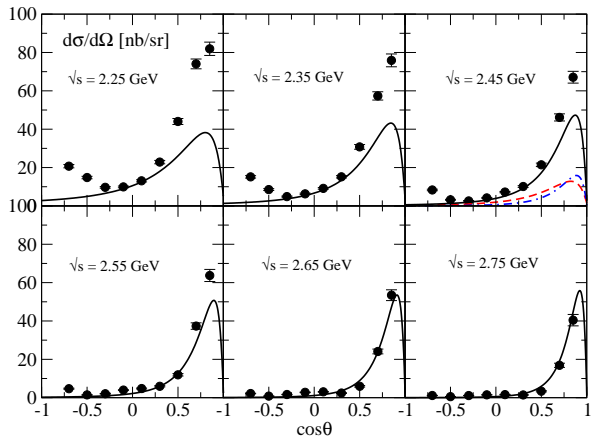} \hspace{0.5cm}
\caption{Differential cross section $d\sigma/d\Omega$ versus
$\cos\theta$ for $\gamma p\to p\eta'(958)$. The solid curve
results from $\rho+\omega$ Reggeon exchanges with a factor of
$0.43$ taken to account for the branching fraction from the
$\gamma p\to p\eta\pi^+\pi^-$ experiment. Below $\sqrt{s}\approx
2.55$ GeV there is a room for nucleon resonances to contribute to
the backward rise as well as the forward enhancement. Notations
for $\rho$ and $\omega$ contributions are the same as in Fig.
\ref{fig2}. Data are taken from Ref.~\cite{dickson}. \\
}
\label{fig3}
\end{figure}
\begin{figure}[]
\centering \epsfig{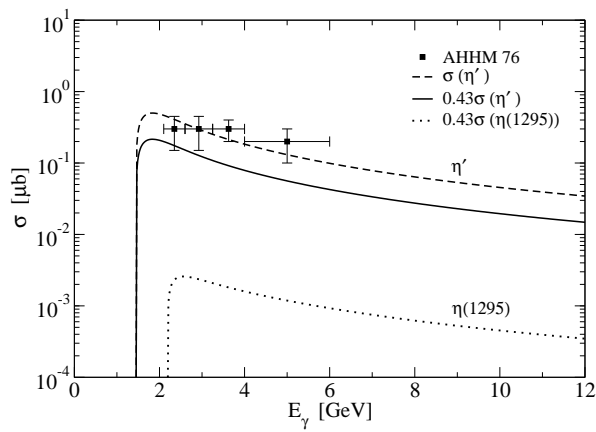} \caption{Total
cross sections for $\gamma p\to p\eta'$ and $\gamma p\to
p\eta(1295)$. The solid and dotted curves correspond to $\eta'$
and $\eta(1295)$ production cross sections scaled by the branching
ratio. For comparison, the dashed curve for $\eta'$ cross section
without scaling is presented to reproduce the
AHHM data \cite{struc76}.} \label{fig4}
\end{figure}

In the next we calculate the exclusive $\gamma p\to p\eta'$ and
$\eta(1295)$ within the same framework. However, in order to
compare with the CLAS data from the multimeson photoproduction
process $\gamma p\to\eta(958)p\to
p\eta\pi^+\pi^-$, we consider the branching ratio
$\Gamma_{\eta'\to\eta\pi^+\pi^-}/\Gamma_{\eta'\to all}\approx
43\%$, from PDG to implement the reduction by an overall factor
of $0.43$ to the cross section from the exclusive process
$\gamma p\to p\eta'$. In Table
\ref{tb1} we list the coupling constants compiled for the $\eta'$
and $\eta(1295)$ in addition to $\eta$ photoproduction above. For
the physical status of $\eta(1295)$ only the mass
$m_{\eta(1295)}=1294\pm4$ MeV and decay width
$\Gamma_{\eta(1295)}=55\pm5$ MeV are known. We follow the
coupling constants deduced from Ref. \cite{kochelev} to
calculate total cross section with the same ratio of the reduction
as in the case of $\eta'$.

Figures \ref{fig3} and \ref{fig4} show the differential and total
cross section for $\eta'$ and total cross section for $\eta(1295)$.
As before, the complex phase $e^{-i\pi\alpha}$ is
chosen for both $\rho+\omega$ Reggeons in both reactions.
Data of differential cross sections are from
the CLAS measurement and the data for the total cross section in
Fig. \ref{fig4} are from the AHHM \cite{struc76}.
These reaction cross sections are scaled,  as discussed above.
The angular distribution reproduced in Fig. \ref{fig3}
is consistent with data.
Nevertheless, in addition to the $t$-channel exchanges,
the underestimate of differential data below
$\sqrt{s}\approx 2.55$
GeV implies the need for the contribution of baryon resonances in the
$s$- and $u$-channel in order to reproduce
the backward rise as well as the forward enhancement in the cross section.
As for the total cross sections in Fig. \ref{fig4}, we first note
that the model prediction without correction
is consistent with the exclusive AHHM cross section \cite{struc76}
as shown by the dashed curve. Within the present framework the
$\eta(1295)$ cross section is smaller than the $\eta'$ by two
orders of magnitude, and, hence, it is reasonable to neglect
the $\eta(1295)$ production in the data analysis for
$f_1$ production from the reaction $\gamma p\to p\eta\pi^+\pi^-$,
as performed by the CLAS Collaboration.

\section{$f_1(1285)$ photoproduction on the proton target}

\subsection{Analysis of CLAS data from $\gamma p\to pf_1\to p\eta\pi^+\pi^-$
below $\sqrt{s}=3$ GeV}

In the CLAS experiment on
the $\gamma p\to p\eta\pi^+\pi^-$ reaction,
the structure of $f_1(1285)$ was observed at $m_X\simeq1280$ MeV of
the $\gamma p$
missing mass spectrum with a great
statistics $\simeq 1.5 \times 10^5\times(1280)$ events. Since
$\eta(1295)$ as well as $f_1(1285)$ is decaying to $\eta\pi\pi$,
care must be taken for the potential overlap with each other to
extract the structure associated with the $f_1(1285)$ with
$p$-wave decay and positive parity from the Dalitz analysis of
$x\to \eta\pi^+\pi^-$. The experiment leads to a conclusion
on $f_1$ with mass $m_{f_1}=1281.0 \pm 0.8$ MeV and width
$\Gamma_{f_1}=18.4\pm 1.4$ MeV which is narrower than PDG value $24.2 \pm
1.1$ MeV.

In the theoretical side, Kochelev \cite{kochelev} and Domokos
\cite{domokos} calculated the reaction cross section for the
exclusive $\gamma p\to pf_1$ by using the Reggeized model for
the $t$-channel $\rho$ and $\omega$ vector meson exchanges.
As depicted in Fig. \ref{fig1} where the outgoing $\eta$ meson
is now replaced by the $f_1$ meson in the Feynman diagram,
the following form of the $\gamma VA$
coupling vertex,
\begin{eqnarray}
\Gamma^\beta_{\gamma VA}(k,Q)\eta_\beta={g_{\gamma VA}\over m_0^2}
Q^2\epsilon^{\mu\nu\alpha\beta}\epsilon_\mu
k_\nu\xi_{\alpha}^*\eta_\beta, \hspace{0.5cm}\label{gva1}
\end{eqnarray}
is utilized in Ref. \cite{kochelev} for the $t$-channel
vector meson exchange. Here $\xi^\alpha(q)$ and $\eta^\beta(Q)$
are spin polarizations of axial vector meson and vector meson with
the momenta $q$ and $Q$, respectively.
Given the $VNN$ coupling vertex  in Eq. (\ref{vnn}),
the vector meson exchange is now written as
\begin{eqnarray}\label{amp2}
&&\mathcal{M}_V=
\Gamma^\beta_{\gamma VA}(k,Q)(-g_{\beta\lambda}+Q_\beta Q_\lambda/m^2_V)
\nonumber\\&&\times\overline{u}(p')\left(g^v_{VNN}\gamma^\lambda+{g^t_{VNN}\over
4M}[\gamma^\lambda, \rlap{/}Q]\right)u(p){\cal R}^V(s,t)
\end{eqnarray}
with the hadron form factors of the form,
\begin{eqnarray}\label{ff}
\left({\Lambda_1^2-m_V^2\over \Lambda_1^2-t}\right)\, {\rm and}\,
\left({\Lambda_2^2-m_V^2\over \Lambda_2^2-t}\right)^n,
\end{eqnarray}
which are assigned to the $\gamma Vf_1$
and $f_1NN$ vertices, respectively, in the amplitude in Eq. (\ref{amp2}).

The decay width of the $\gamma VA$ vertex corresponding to Eq. (\ref{gva1}) is
given by
\begin{eqnarray}\label{gva-decay}
\Gamma_{A\to V\gamma}={1\over 96\pi}{g^2_{\gamma VA}\over
m_0^4}{m_V^2\over m_A^5} (m_A^2+m_V^2)(m_A^2-m_V^2)^3,\hspace{0.5cm}
\end{eqnarray}
and the coupling constant $g_{\gamma\rho^0f_1}=0.94$ is
determined from the  decay width $\Gamma_{f_1\to\rho^0\gamma}\approx1330$ keV
presently reported in the PDG. The coupling constant
$g_{\gamma\omega f_1}=-g_{\gamma\rho f_1}/3$ is taken from the
quark model estimation.

However, with the trajectories and $VNN$ coupling constants given
in Ref. \cite{kochelev}, the model prediction for the CLAS
data is poor, even though the model
assumes the reduction of the calculated cross section
by the branching ration 35$\%$ as applied in
Figs. \ref{fig3} and \ref{fig4} for comparison with the CLAS
cross section $\gamma p\to pf_1\to p\eta\pi^+\pi^-$.
The crucial point of the issue is that the PDG width $f_1\to\rho^0\gamma$
chosen above
is too large to agree with the CLAS data.
Theoretical estimates based on the QCD inspired models such as the
constituent quark model (CQM) \cite{ishida} and the Four quark state
with the triangle loop for the $AVV$ anomaly \cite{osipov}
suggest half the value of the current PDG fit. Moreover, the
width $453\pm 177$ keV extracted from the CLAS experiment further supports
these smaller values rather than the PDG one as shown in Table \ref{tb2}.

\begin{table}
\centering\caption{Estimate of $\gamma VA$ coupling constant
from the decay width of Ref. \cite{dickson}$^a$,
Ref. \cite{ishida}$^b$, Ref. \cite{osipov}, and PDG
which are given in units of keV. For comparison we list $g_{\gamma\rho
f_1}=0.59^b$, $0.45^c$ and $0.94^d$. $g_{\gamma\omega
f_1}=0.152^c$ from these references.
}
\begin{tabular}{ccccccc}
\hline\hline
                       &$g_{\gamma VA}$&CLAS\cite{dickson}&CQM\cite{ishida}&4quark\cite{osipov} &PDG  &\\%
\hline
$f_1\to\rho\gamma$     &  $0.54^a$     & 453$^a$          & $509^b$      &$311^c$    &$1330^d$ &\\%
$f_1\to\omega\gamma$   & $-0.18^b$      &  -               & $48^b$       &$34.3^c$    &-   & \\%
\hline\hline
\\
\end{tabular}
\label{tb2}
\end{table}

Avoiding the model dependence such as the cutoff mass with form
factors in Eq. (\ref{ff}), we perform the analysis of the CLAS
data with the decay width $\Gamma_{f_1\to \rho^0\gamma}=453$ keV
determined from the CLAS experiment. We then demonstrate how the
production mechanism could account for the CLAS data, while
comparing our results with Ref. \cite{kochelev}. It is legitimate
to consider simply the $\rho^0+\omega$ exchanges as in Fig.
\ref{fig1}, because threshold energy of the reaction,
$\sqrt{s}_{thres.}\approx2.2$ GeV is high enough to neglect
nucleon resonances. This might be a contradiction to the finding
in Ref. \cite{dickson} that the production mechanism is more
consistent with $s$-channel decay of a high-mass $N^*$ state not
with $t$-channel meson exchange, because the hadron models
aforementioned are insufficient to reproduce the CLAS data. The
role of nucleon resonance $N^*(2300)(1/2^+)$ together with the
nucleon in the $s$ and $u$-channel is discussed to account for the
$u$-channel rise at $\sqrt{s}=2.65$ and 2.75. \cite{wang}.
Nevertheless, it should be pointed out that the results from these
models follow the dependence on the cutoff mass of the form
factors, which is the point that is quite different from the current
calculation.

In the Table \ref{tb2} we choose $g_{\gamma\rho f_1}=0.54$ from the CLAS width
and $g_{\gamma\omega f_1}=-0.18$  from the Relativistic quark
model which resumes the ratio of $g_{\gamma\rho^0f_1}/g_{\gamma\omega f_1}\approx-3$.

\begin{figure}[]
\centering \epsfig{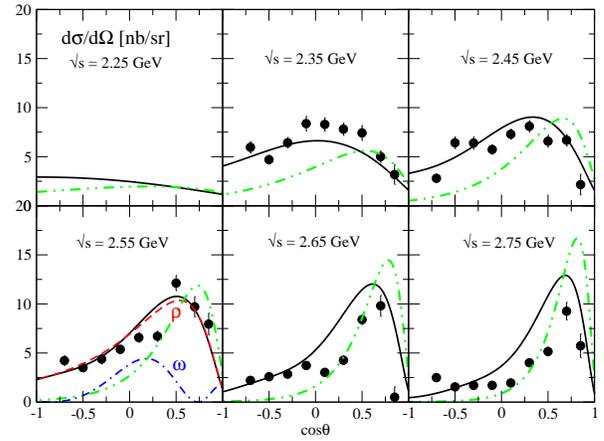}
\caption{Differential cross section $d\sigma/d\Omega$ versus
$\cos\theta$ for $\gamma p\to pf_1\to p\eta\pi^+\pi^-$. The solid
cross sections are scaled by a factor of $0.35$ to account for the
branching fraction from the $\gamma p\to p\eta\pi^+\pi^-$. The
(Green) dash-dot-dotted curve is from Ref. \cite{kochelev} with
$\Lambda_1=1.2$ and $\Lambda_2=1.4$ GeV and $n=1$ for the $\gamma
Vf_1$ form factor. Data are taken from Ref.~\cite{dickson}. \\
}
\label{fig5}
\end{figure}
\begin{figure}[]
\centering \epsfig{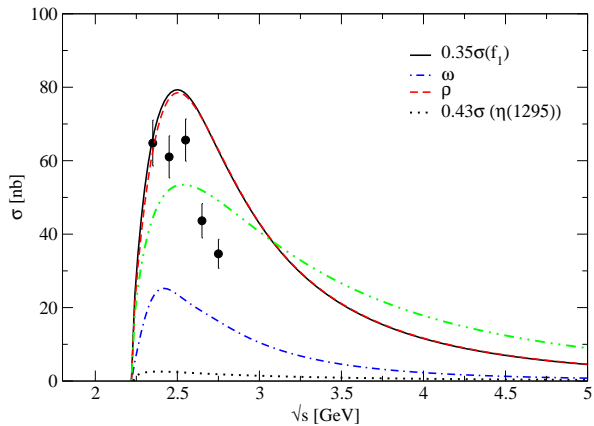}
\caption{Total
cross sections for $\gamma p\to Pf_1\to p\eta\pi^+\pi^-$ and
$\gamma p\to p\eta(1295)\to p\eta\pi^+\pi^-$. Five data points in
the $f_1$ cross section are obtained by integrating out the
differential cross sections given in Fig. \ref{fig5} for
illustration purposes. The cross sections are scaled by the same
factors as in Figs. \ref{fig4} and \ref{fig5}, respectively. For
comparison the cross section for $f_1$ from Ref. \cite{kochelev}
is presented by the green dash-dot-dotted curve. The cross section
for $\gamma p\to\eta(1295)p$ scaled by 43$\%$ is shown to be
negligible in the region overlapping  with $f_1$ production.
  } \label{fig6}
\end{figure}

With the $VNN$ coupling constants in Table \ref{tb1} and the
vector meson trajectories in Eqs. (\ref{trajectory}) and
(\ref{omega-traj}), we reproduce the CLAS differential cross
section in Fig. \ref{fig5}. In order to describe the exclusive
$\gamma p\to pf_1$ reaction from the CLAS data which are extracted
from the $\gamma p\to p\eta\pi^+\pi^-$ in the final state, we have
to consider a scaling of the cross section by the fraction
$\Gamma_{f_1\to\eta\pi^+\pi^-}/\Gamma_{f_1\to all}\approx0.35$,
similar to the case of $\eta'$ photoproduction, as before. As the
differential cross section data show no oscillatory behavior, the
complex phase $e^{-i\pi\alpha_{\rho}(t)}$ for the $\rho$ is
mandatory and the canonical phase
$1/2(-1+e^{-i\pi\alpha_{\omega}(t)})$ is chosen for $\omega$  to
be consistent with data. The roles of $\rho$ and $\omega$ are
displayed at $\sqrt{s}=2.55$ GeV.
The (green) dash-dotted curve results from Ref. \cite{kochelev}
with the cutoff masses $\Lambda_1=1.2$
and $\Lambda_2=1.4$ GeV together with $n=1$ for the form factors.
In the model only $g_{\gamma Vf_1}$ is taken the same as
ours for comparison.
It is shown that the cross section is too
much suppressed in the nonforward direction due to the
form factors, which is different from the present approach
without form factors.

\begin{figure}[]
\vspace{.5cm} \centering \epsfig{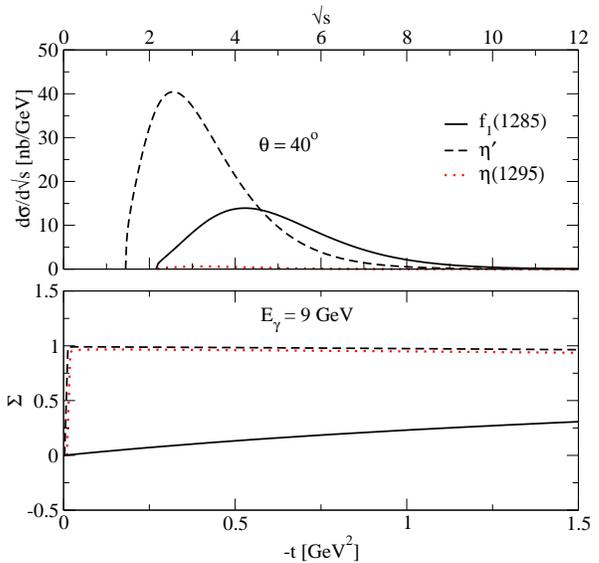}
\caption{Differential cross section and beam polarization
asymmetry for $f_1$, $\eta'$, and $\eta(1295)$ photoproductions
from the CLAS experiment $\gamma p \to p\eta\pi^+\pi^-$.
Differential cross sections scaled by the respective factors are
predicted as a function of $\sqrt{s}$ at the production angle
$\theta^*=40^\circ$ in the center of mass system. The
$t$-dependence of  beam polarizations are presented at
$E_\gamma=9$ GeV with the same notation as in the differential
cross sections. } \label{fig7}
\end{figure}

Total cross sections for $f_1$ and $\eta(1295)$ photoproductions
are shown in Fig. \ref{fig6}. For reference purposes, we present
the $f_1$ cross section data by integrating out the differential
cross section data in Fig. \ref{fig5} over the angle. The
respective contributions of $\rho$ and $\omega$ exchanges are
shown with the same notations. The dash-dot-dotted curve from Ref.
\cite{kochelev}, as in Fig. \ref{fig5}, yields the smaller cross
section at low energy. The cross section for $\eta(1295)$
photoproduction is shown with a correction by the factor of
$0.43$. By comparison we agree with the negligence of $\eta(1295)$
component in the CLAS analysis of $f_1(1285)$ cross section from
the reaction $\gamma p\to p\eta\pi^+\pi^-$.

For further studies on the CLAS data, we present the energy
dependence of differential cross section at the forward angle
$\theta=40^\circ$ (corresponding to a forward peak at
$\cos\theta\approx0.75$ and $\sqrt{s}=2.75$ GeV in Fig.
\ref{fig5}) in Fig. \ref{fig7} for $f_1(1285)$, $\eta(958)$, and
$\eta(1295)$ cross sections involved in the overlapped potential
region. The discrimination between these reactions is more
apparent in the beam polarization $\Sigma$ as can be seen in Fig.
\ref{fig7}. As the present model includes only the natural parity
exchange of $\rho$ and $\omega$, the $\Sigma$ is always positive.
But, the different size of the $\Sigma$ between pseudoscalar and
axial vector meson photoproductions reveals the different scheme
of the interference between $\rho$ and $\omega$ Reggeons among
these reactions.

\subsection{Primakoff effect in the $f_1$ process}

Now that the decay $f_1\to \gamma^*\gamma$ is another source of
the reaction $\gamma p\to pf_1$ to proceed in the forward direction,
the virtual photon exchange serves to the exclusive $f_1$
photoproduction via one of the two photons off-mass shell in the
$t$-channel. This is the so-called the Primakoff effect
which is observed at very forward angles, in general, in the photoproduction
of charge-neutral meson of $C$-parity even \cite{kaskulov,sibirtsev}.
The Primakoff effect provides an opportunity to test non-perturbative
properties of QCD through the mixing of the nonet members in the
axial anomaly \cite{gasparian}.
Moreover, as the virtual photon exchange is not to be reggeized,
it is not subject to the energy dependence $s^{\alpha(0)-1}$ at high energies.
Therefore, by virtue of it, such energy independent behavior
is expected in the reaction, as in the case of the Pomeron exchange
at high energies in vector meson photoproduction.
The nature of the Pomeron and virtual photon exchange is,
of course, quite different. Because the origin of the
former process comes from strong interaction by the exchange of two
gluon correlation \cite{landshoff}, whereas
the latter exchange results from the axial anomaly in
the presence of electromagnetic interaction.
Thus, this subprocess in the $t$-channel will be an example
of seeking the nonmesonic process that can
survive at high energies in the photoproduction
of axial vector meson.

Given the $\gamma VA$ vertex in Eq. (\ref{gva1}),
we replace the vector meson polarization $\eta_\beta$
by the virtual photon $\epsilon'_\beta$ with 4-momentum $Q^\mu$ in
the $t$-channel for the virtual photon exchange.
Then, the $\gamma^*$ exchange is now written as
\begin{eqnarray}\label{2photon}
&&\mathcal{M}_{\gamma^*}=
-\Gamma^\beta_{\gamma\gamma^* A}(k,Q)F_{\rho}(t)
{\left(-g_{\beta\lambda}\right)\over t}
\nonumber\\&&\hspace{1cm}
\times e\,\overline{u}(p')\left(e_N\gamma^\lambda
+{\kappa_{N}\over4M}[\gamma^\lambda, \rlap{\,/}Q]\right)F_{1}(t)u(p)\,,
\end{eqnarray}
where the coupling constant $g_{\gamma VA}$ in Eq. (\ref{gva1})
is read as $g_{\gamma\gamma^*A}$ and the vector meson polarization
$\eta$ as $\epsilon$ with the vector meson propagator and $VNN$ vertex
replaced by the virtual photon propagator and $\gamma NN$ vertex
in Eq. (\ref{amp2}). $e_N=1$ for proton with $\kappa_p=1.79$ and 0
for neutron with $\kappa_n=-1.91$.
The coupling constant $g_{\gamma\gamma^*A}$ now in Eqs. (\ref{gva1})
and (\ref{2photon})
cannot be estimated from the decay width as before, because of
vanishing of $Q^2$ for the case of real photons.
Following Ref. \cite{osipov}, we determine the coupling constant
$g_{\gamma\gamma^* A}$  in Eq. (\ref{gva1}) as
\begin{eqnarray}
{g_{\gamma\gamma^* A}\over m_0^2}=8\pi\alpha
F^{(0)}_{AV\gamma^*\gamma^*}(m_f^2,0,0),
\end{eqnarray}
which yields $g_{\gamma\gamma^*f_1}$=0.043 by taking
$F^{(0)}_{AV\gamma^*\gamma^*}(m_f^2,0,0)=(0.234\pm0.034)$ GeV$^{-2}$
from the PDG.

Since we are dealing with the virtual photon exchange,
the following vertex form factors
\begin{eqnarray}
&&F_{\rho}(t)=\left(1-t/m_\rho^2\right)^{-1},
\label{vmd}\\
&&F_{1}(t)={4M^2-2.8t\over(4M^2-t)(1-t/0.71\,{\rm GeV}^2)^2}
\label{f1nn}
\end{eqnarray}
are introduced to $\gamma\gamma^* f_1$
and $\gamma^*NN$ vertices  in the $t$-channel.
By the vector meson dominance, a combination of
$\rho$-$\omega$-$\phi$ meson poles should be applied to the
$\gamma\gamma^*f_1$ vertex
with the mixing between $\rho+\omega$ and $\phi$
vector mesons
as discussed in Ref. \cite{kaskulov} for the cases of $\gamma\gamma^*\eta$
and $\gamma\gamma^*\eta'$ vertices.
Suppose that there is no difference between  vector
meson masses, then, these form factors lead roughly to a unity
as the mixing has no meaning between $\rho+\omega$
and $\phi$ \cite{kaskulov}. Thus, we choose the
$\rho$-meson pole here as a representative for simplicity.
For the $\gamma^*NN$ vertex, we employ the nucleon isoscalar form factor
in Eq. (\ref{f1nn}) which replaces the Dirac form factors
$F_1$ and $F_2$, as discussed in Ref. \cite{donnachie}.
The form factors in Eqs. (\ref{vmd}) and (\ref{f1nn}) with
cutoff masses are well established in other hadronic processes so that
we have no model dependence in calculating
reaction cross sections for the $\gamma p\to pf_1$ reaction.

\begin{figure}[]
\centering \epsfig{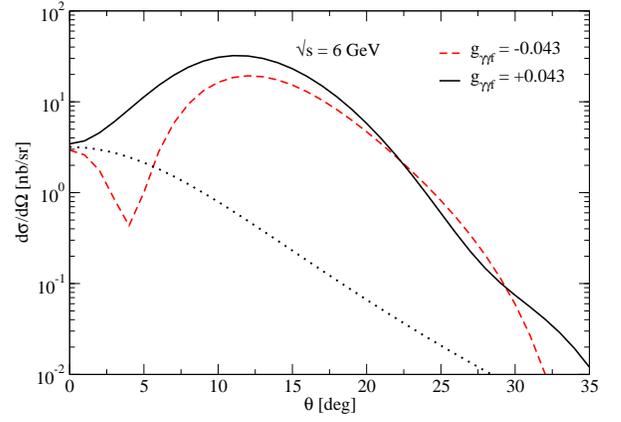}
\caption{Differential cross section for the exclusive $\gamma p\to
pf_1$ reaction as a function of angle $\theta$. The Primakoff
effect between positive and negative sign of the coupling constant
$f_{\gamma\gamma^*f_1}$ shows a different angle dependence below
$\theta\approx 5^\circ$. \\
\\
 }\label{fig8}
\end{figure}

\begin{figure}[]
\centering \epsfig{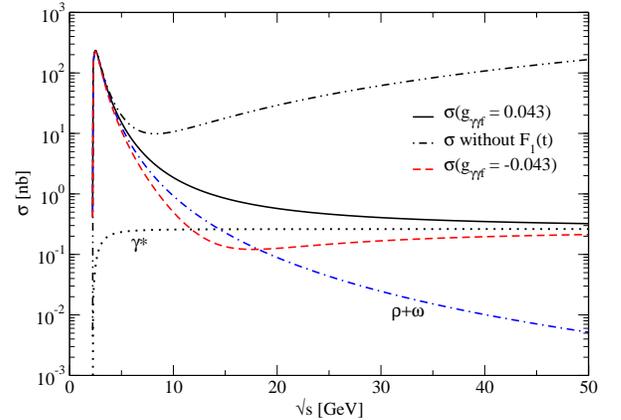} \caption{Total
cross section for the exclusive $\gamma p\to pf_1$ process up to
$\sqrt{s}=50$ GeV. Contributions of $\gamma^*$ and $\rho+\omega$
exchanges are shown by the dotted and dash-dotted curves,
respectively. The solid (dashed) curve corresponds to total cross
section with the relative sign of $\gamma^*$ exchange positive
(negative) to vector meson exchanges. The exchange of $\gamma^*$
gives the contribution $\sigma\approx 0.25$ nb at $\sqrt{s}\approx
50$ GeV persistent up to the higher energies. The solid cross
section without form factor $F_1(t)$ diverges, as shown by the
dash-dot-dotted curve. }\label{fig9}
\end{figure}

In Fig. \ref{fig8} differential
cross section is presented at forward angles below $\theta=35^\circ$
at $\sqrt{s}=6$ GeV and total cross section is shown up to
$\sqrt{s}\approx50$ GeV in Fig. \ref{fig9}.
Since there is no information about
the sign of the $\gamma\gamma^*f_1$ coupling relative to that of the vector
meson exchange, $\gamma V f_1$, the cross section is shown for
both signs of $g_{\gamma\gamma^*f_1}$ for illustration purposes.
The solid and dashed (red) curves are from the full calculation
of the differential and total cross sections corresponding to positive
and negative signs. The exchange of $\gamma^*$ is denoted by the dotted
curve in both cross sections.


It is found that in the total cross section the role of form
factor $F_{\rho}(t)$ in Eq. (\ref{vmd}) is negligible. However,
without the nucleon isoscalar form factor $F_1(t)$, the cross
section is highly divergent, as can be seen by the dash-dotted
curve.  The energy dependence of the cross section shows a growth
of the Primakoff effect up to $\sqrt{s}\approx5$ GeV and remains
constant, i.e., $\sigma\approx0.1$ nb up to $\sqrt{s}\approx50$
GeV. Beyond $\sqrt{s}\approx25$ GeV the contribution of $\gamma^*$
exchange becomes stronger than those of vector meson exchanges
which are decreasing by the energy dependence $\sim
s^{\alpha_V(0)-1}$.

\subsection{Primakoff effect in pseudoscalar meson photoproduction}

Pseudoscalar meson photoproduction is a typical reaction to
observe the Primakoff effect by the $\gamma^*$ exchange
\cite{kaskulov}\cite{sibirtsev}.
A precise measurement of the Primakoff effect has been advocated
to test the nonperturbative QCD based on the flavor mixing of the chiral
symmetry \cite{gasparian}.
Here we shall reproduce the Primakoff effect in the photoproduction
of $\pi^0$ and $\eta$ as well as $\eta'$ in the PrimEx energy region
\cite{gasparian} and compare the result with the data available.
Nevertheless, our interest in this issue is still more in understanding the
role of the nonmesonic scattering in the pseudoscalar meson
photoproduction at high energies, as demonstrated in the $f_1$
photoproduction.

For the Primakoff effect the virtual photon exchange in the
$\eta$ photoproduction is written as
\begin{eqnarray}\label{prima3}
&&{\cal M}_{\gamma^*}={f_{\gamma\gamma^*\eta}\over m_\eta}
F_{\gamma\gamma^*\eta}(t)
\epsilon^{\mu\nu\alpha\beta}\epsilon_\mu k_\nu Q_\alpha
{\left(-g_{\beta\lambda}\right)\over t}
\nonumber\\&&\hspace{1cm}
\times e\,\bar{u}(p')\left(e_{N}\gamma^\lambda+{\kappa_N\over
4M}[\gamma^\lambda, \rlap{/}Q]\right)F_{1}(t)u(p),
\end{eqnarray}
where
the vertex form factors $F_{\gamma\gamma^*\eta}(t)$
and $F_1(t)$ are given by  Eqs. (\ref{vmd}) and (\ref{f1nn}),
respectively.
The radiative coupling constant $f_{\gamma\gamma^*\eta}$
is estimated from the decay width
\begin{eqnarray}
\Gamma_{\eta\to\gamma\gamma}={f_{\gamma\gamma\eta}^2m_\eta
\over64\pi}
\end{eqnarray}
with the PDG value taken for the $\eta\to\gamma\gamma$ decay. 

\begin{figure}[]
\centering \epsfig{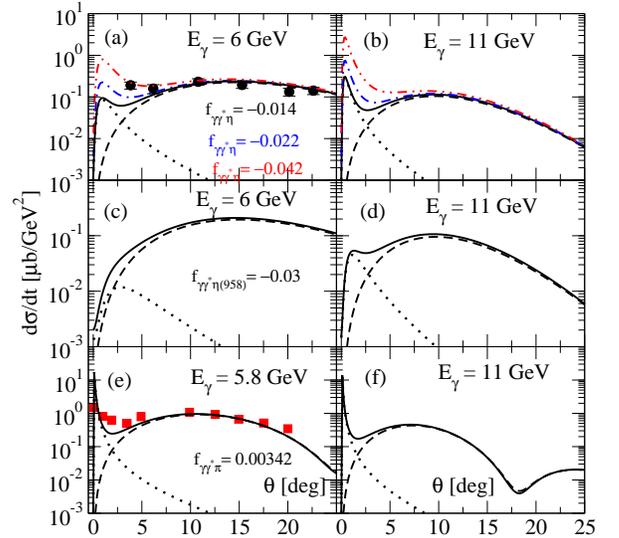}
\caption{Differential cross sections for $\gamma p\to p\eta$
in (a) and (b),
$\gamma p\to p\eta'$ in (c) and (d),
and $\gamma p\to p\pi^0$ in (e) and (f).
The (red) dash-dot-dotted, (blue) dash-dotted, and
solid curves result from the case of coupling constant
$f_{\gamma\gamma^*\eta}$ as denoted in the panel (a),
respectively.
In (f) the dip at $\theta\approx18^\circ$ is due to the
nonsense zero of $\omega$ exchange.
Data of $\eta$ and of $\pi^0$ production are taken from
Ref. \cite{braunschweig70} and
Ref. \cite{mbraunschweig70}, respectively.
}\label{fig10}
\end{figure}

In Fig. \ref{fig10} the differential cross sections for $\gamma
p\to p\eta$ with the DESY data \cite{braunschweig70}, and $\gamma
p\to p\eta'$ and $\gamma p\to p\pi^0$ \cite{mbraunschweig70} are
presented to exhibit the role of $\gamma^*$ exchange at very
forward angles. The solid curve results from the
$\rho+\omega+\gamma^*$ exchanges in Eqs. (\ref{vec}) and
(\ref{prima3}) with $f_{\gamma\gamma^*\eta}/m_\eta=-0.014/m_\eta$
from $\Gamma_{\eta\to\gamma\gamma}=0.52$ keV,
$f_{\gamma\gamma^*\eta'}/m_{\eta'}=-0.03/m_{\eta'}$ from
$\Gamma_{\eta'\to\gamma\gamma}=4.28$ keV and
$f_{\gamma\gamma^*\pi^0}/m_{\pi^0}=+0.00342/m_{\pi^0}$ from
$\Gamma_{\pi^0\to\gamma\gamma}=7.74$ eV cases, respectively
\cite{amsler}. The coupling constants $g_{\gamma\pi^0\rho}=0.255$
and $g_{\gamma\pi^0\omega}=0.7$ are used for the $\pi^0$ cross
sections with the $VNN$ coupling constants given in Table
\ref{tb1}. The trajectories are taken the same as the $\eta$ case,
but the phase $e^{-i\pi\alpha_\rho}$  and
${1\over2}(-1+e^{-i\pi\alpha_\omega})$ are chosen for $\rho^0$ and
$\omega$, respectively.
In the case of $\eta$, the negative sign is
chosen for the constructive interference between $\gamma^*$
and $\rho+\omega$ exchanges.
However, the result shows a discrepancy
with experimental data, in particular,
below $\theta\approx5^\circ$, which is comparable to that of
Ref. \cite{kaskulov} at the same energy.
In order to agree with the data the coupling constant $f_{\gamma\gamma^*\eta}$
should grow by three times larger than the one given above,
as shown by the (red) dash-dot-dotted curve.
In this case the result is similar to Ref. \cite{sibirtsev}.
This is, however, unattainable, even though the maximum value
for the mixing angle is chosen between $\eta_0$ and $\eta_8$,
when we express the decay width $\eta\to\gamma\gamma$ and
$\eta'\to\gamma\gamma$ as in Eqs. (\ref{width}) and (\ref{width1})
given below.

In consideration of the mixing of the $SU_f(3)$ flavor singlet and
octet members the decay widths for $\eta\to\gamma\gamma$ and
$\eta'\to\gamma\gamma$ are written as \cite{gasparian}
\begin{eqnarray}\label{width}
&&\Gamma_{\eta\to\gamma\gamma}={\alpha^2\over64\pi^3}
{m_\eta^3\over3f_\pi^2}\left({f_\pi\over f_{\eta_8}}\cos{\theta_p}
-\sqrt{8}{f_\pi\over f_{\eta_0}}\sin{\theta_p}\right)^2,\hspace{0.5cm}\\
&&\Gamma_{\eta'\to\gamma\gamma}={\alpha^2\over64\pi^3}
{m_{\eta'}^3\over3f_\pi^2}\left({f_\pi\over f_{\eta_8}}\sin{\theta_p}
+\sqrt{8}{f_\pi\over f_{\eta_0}}\cos{\theta_p}\right)^2,\hspace{0.5cm}
\label{width1}
\end{eqnarray}
where $f_{\eta_0}$ and $f_{\eta_8}$ are the flavor singlet
and octet decay constants and $\theta_p$ is the mixing angle
for pseudoscalar mesons. They are estimated as
$f_{\eta_8}\approx 1.3f_\pi$ by ChPT and $f_{\eta_0}\approx f_\pi$
in the Large $N_c$ limit. The ratio measured in the experiment
\begin{eqnarray}
R={1\over3}\left({f_\pi^2\over f^2_{\eta_8}}+8{f_\pi^2\over f^2_{\eta_0}}\right)
=2.5\pm0.5
\end{eqnarray}
is consistent with the theoretical estimates,
$f_\pi/f_{\eta_0}=0.93$ with $f_\pi/f_{\eta_8}=1/1.3$.

Let us now make an estimate of the decay width
based on Eqs. (\ref{width}) and (\ref{width1}).
In terms of $f_{\eta_0}=100.1$ MeV and $f_{\eta_8}=121$ MeV 
that are given by taking $f_\pi=93.1$ MeV above,
the $\eta$ decay width $0.52$ keV taken here corresponds to the mixing
angle $\theta_p\approx-23.4^\circ$. At the
angle the corresponding $\eta'$  decay width leads to 4.03 keV
from Eq. (\ref{width1}), which is close to the empirical value quoted above.
In practice the maximum angle $\theta_p$
in Eq. (\ref{width}) exists at $-74^\circ$ (or $106^\circ$),
which yields the decay width $1.27$ keV.
The (blue) dash-dotted curve corresponds to the decay
width at such an angle $\theta_p$ with $f_{\gamma\gamma^*\eta}=-0.022$,
which is, however, still deficient to agree with data. Moreover,
in that case the $\eta'$ decay width from Eq. (\ref{width1}) is
vanishing, i.e about 0.2 eV at the angle.
Therefore, from the relations between $\eta$ and
$\eta'$ mixing above, the coupling constant $f_{\gamma\gamma^*\eta}=|0.042|$
cannot be achievable and we notice that the discrepancy below the production angle
$\theta\approx 5^\circ$ can no longer be covered over even with
the mixing angle maximally allowed. In future experiments such as
the PrimEX project at CLAS 12 GeV a precise measurement of
the cross section at very forward angles is desirable to decide
whether such a disagreement is still due to a theoretical deficiency,
or an experimental uncertainty.

\begin{figure}[]
\centering \epsfig{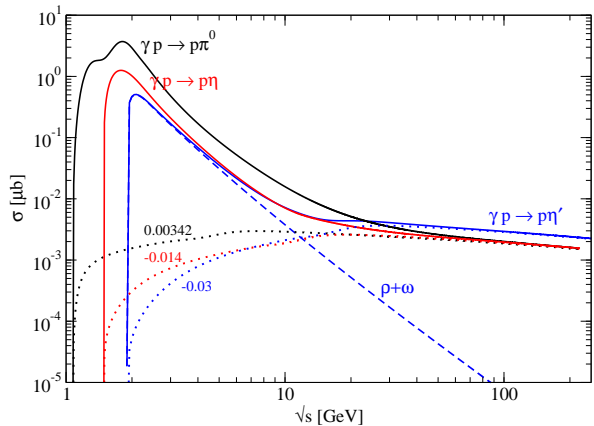}
\caption{Primakoff effect in $\gamma p\to p\pi^0$, $\gamma p\to
p\eta$ and $\gamma p\to p\eta'$ reactions at high energies. The
dotted curve is the contribution of $\gamma^*$ exchange which
shows a nearly energy independent behavior up to
$\sqrt{s}\approx250$ GeV. The rapid peak of the total cross
section near threshold is reproduced by the $\rho+\omega$
exchanges depicted by the dashed curve in the $\eta'$ process, for
instance. }\label{fig11}
\end{figure}

Figure \ref{fig11} shows total cross sections for $\gamma p\to
p\pi^0$, $\gamma p\to p\eta$, and $\gamma p\to p\eta'$ reactions
up to $\sqrt{s}\approx 250$ GeV. The solid cross sections are from
the $\rho+\omega+\gamma^*$ exchanges in the $t$-channel. Beyond
$\sqrt{s}\approx100$ GeV where there exists the dominating
$\gamma^*$ exchange, the $\pi^0$ and $\eta$ cross sections are
persistently scaled to a common limit $\sigma\simeq1.5$ nb around
$\sqrt{s}\simeq250$ GeV. This coincidence is understood, if one
notes that $f_{\gamma\gamma^*\pi^0}\approx f_{\gamma\gamma^*\eta}$
in unit of GeV$^{-1}$. The $\eta'$ cross section reaches at the
limit $\sigma\simeq2.2$ nb with the coupling constant larger than
the two. Together with the  $f_1$ case as shown in Fig.
\ref{fig9}, the role of $\gamma^*$ exchange in these cross
sections scaling up to 250 GeV can be compared to that of the
Pomeron exchange in the vector meson photoproduction. The limit of
the cross section at high energies is determined only by the
coupling strength of each meson decaying to two gamma's.
Therefore, a measurement of the $\eta$ and $\eta'$ cross sections
at such limiting energies enables us to determine the mixing
angles from Eqs. (\ref{width}) and (\ref{width1}).

Before closing this section, we would give a remark on the
specialty of photoproductions of pseudoscalar and axial vector
meson, which exhibits the dynamical features as interesting as the
Pomeron exchange in the vector meson photoproduction at high
energies. Of course, in these reactions besides the virtual photon
exchange we note that there is an important mechanism to prove
QCD, called the exchange of Odderon which can be another entity
comparable to the Pomeron \cite{berger,adloff}. Likewise the
Pomeron exchange in the high energy vector meson photoproduction,
the search of the Odderon is a most interesting topic to study the
role of odd numbers of gluons in hadron reactions. Thus, to
predict the $\gamma^*$ exchange as in Figs .\ref{fig9} and
\ref{fig11} should be important to find out the Odderon exchange
in future experiments, if available at such high energies. This
topic will be our next work which appears elsewhere.

\section{summary}

The first half of the present work is devoted to analyzing CLAS
data on the $\eta'$ and $f_1(1285)$ photoproductions based on the
$\rho+\omega$ Reggeon exchanges.  The observables of $\eta$
photoproduction is reproduced to confirm the validity of the
vector meson contributions, prior to the studies of the $\eta'$
and $\eta(1295)$ photoproductions. To describe the exclusive
reaction $\gamma p\to p\eta'$ from the multi-meson photoproduction
in the final state $\gamma p\to p\eta\pi^+\pi^-$, and similarly in
the case of $\eta(1295)$, the reaction cross section is corrected
by the branching ratio, taking into account the final decay mode
reported in the PDG. The Regge calculation of axial vector meson
$f_1(1285)$ photoproduction is performed within the same
framework. It is found that the $\eta(1295)$ photoproduction is
small enough to be neglected in the reconstruction of $\gamma p
\to pf_1$ from the reaction $\gamma p\to p\eta\pi^+\pi^-$. Our
model could reproduce the differential cross section to a good
degree, if the branching ratio for $f_1\to\eta\pi^+\pi^-$ of
35$\%$ and the decay width 453 keV are feasible to use. To
demonstrate the production mechanism different between
pseudoscalar and axial vector meson photoproduction, predictions
for the energy dependence of differential cross sections and the
$t$-dependence of the beam polarization asymmetry are presented.
In particular, the beam polarization asymmetry shows the feature
quite contrasting to each other.

The rest of the present work is focused on the exclusive
$f_1$ photoproduction with a special interest in the search of
nonmesonic scattering process such as the Pomeron exchange
in the vector meson photoproduction.
The Primakoff effect  by the virtual photon exchange
shows the behavior of energy independence at high energies
so that the total cross section remains constant persistently
up to $\sqrt{s}\approx50$ GeV with the size of
$\sigma\approx 2$ nb. This feature from the virtual photon exchange
is also implemented in the pseudoscalar meson photoproduction
with the cross section approaching to the limiting
value 1.5 nb in the $\pi^0$ and $\eta$, and 2.2 nb in the $\eta'$ cases
at high energies.

These results provide useful information for the study of
the Primakoff effect by the PrimEX project at CLAS 12 GeV,
and also the detailed analysis of $\gamma p\to p\eta\pi^+\pi^-$
reaction presented in this work helps searching for exotic mesons
via multimeson photoproduction in the GlueX
project at the Jefferson Lab.

       \section*{Acknowledgments}
This work was supported by the National Research Foundation of
Korea Grant No. NRF-2017R1A2B4010117.


\end{document}